\documentstyle[12pt]{article}
\def\Tr{{\rm Tr}}

\begin{document}
\begin{center}     
{\LARGE {\bf The Geometric Construction of WZW }}
\vskip 0.2cm
{\LARGE{\bf Effective  Action in  Non-commutative Manifold}}

\vskip 1cm

{\large Boyu Hou, Yongqiang Wang, Zhanying Yang and Ruihong Yue}
{}~\\
\quad \\
{\em The Institute of Modern Physics(IMP),}\\
{\em Northwest University}\\ 
{\em Xi'an 710069 P.R.China}\\
 {\tt byhou,wyq,yzy,yue@phy.nwu.edu.cn}
\end{center}

\vskip 2cm
\noindent
{\bf Abstract.}   By constructing close
one cochain density
${\Omega^1}_{2n}$ in the gauge  group space we get  WZW effective
Lagrangian on high dimensional
non-commutative space.Especially consistent anomalies derived from this WZW
effective action in non-commutative four-dimensional space coincides with
those by L.Bonora etc\cite{Bonora}.

\vskip 1cm

\section {Introduction}
\noindent
The growing interest in non-commutative geometry is calling our attention
upon some
old problems of YM theory in non-commutative manifold. We would like to
know whether or to what extent old
problems, solutions
or algorithms in commutative YM theories fit in the new non-commutative
framework.
One of these is what form the WZW effective action takes in the new
non-commutative
setting. This has partially  been answered via a direct  computation in 
some special conditions in non-commutative low dimension space \cite{TWO}. 
But the generalization to 
high dimensions is very difficult,even impractical. So we try to 
construct the WZW effective action in
non-commutative high dimensions via its geometric property. In fact,in this
letter WZW
effective action appears
as the close-one-cochain of the gauge group in geometry 
\cite{WZ,RS,BC,ZWZ,HOU,Guo},which is invariant and only dependent on
the topology of gauge group instead of the local geometry,for
example,commutator.
In sequence,this makes the calculation of 
the WZW effective action to be possible through the  
close-one-cochain of gauge group in
the non-commutative high dimensions space.

\vspace{10mm}
\section {Non-commutative Gauge Theory and WZW Effective Action}
\noindent 
In the non-commutative geometry formalism, geometric spaces are described
by a
$C^{\star}$-algebra, which is in general not commutative and realized by Moyal product:
\begin{eqnarray}\label{Moyal}
f\left(x\right)\star g\left(x\right)\equiv e^{\frac{i\theta_{\mu\nu}}{2}\
\frac{\partial}
{\partial\xi_{\mu}}\ \frac{\partial}{\partial\zeta_{\nu}}
}f\left(x+\xi\right)
g\left(x+\zeta\right)\bigg|_{\xi=\zeta=0},
\end{eqnarray}
where  $\theta_{\mu\nu}$ is a real antisymmetric constant 
background, and reflects the
non-commutativity of the coordinates of {\bf R}$^D$ \cite{SW}:
\begin{eqnarray}\label{noncom}
[x_{\mu},x_{\nu}]_{\star}=i\theta_{\mu\nu}.
\end{eqnarray}
Then we have the following identities:
\begin{eqnarray}
\int\limits_{-\infty}^{+\infty}d^{D}x\ f\left(x\right)\star
g\left(x\right) =
\int\limits_{-\infty}^{+\infty}d^{D}x\ g\left(x\right)\star
f\left(x\right)=
\int\limits_{-\infty}^{+\infty}d^{D}x\
f\left(x\right)g\left(x\right),\label{com}
\end{eqnarray}
\begin{eqnarray}\label{cycle}
\int\limits_{-\infty}^{+\infty}d^{D}x\ \left(f_{1}\star f_{2}\star
f_{3}\right)
\left(x\right) =
\int\limits_{-\infty}^{+\infty}d^{D}x\ \left(f_{3}\star f_{1}\star
f_{2}\right)
\left(x\right) = \int\limits_{-\infty}^{+\infty}d^{D}x\ \left(f_{2}\star
f_{3}\star f_{1}  
\right)\left(x\right).
\end{eqnarray}
Especially , in integral with one compact space there is the following
circular relation, i.e. the following relation can work up to a space-time
derivative \cite{Bonora}:
 \begin{eqnarray}
\Tr(E_1*E_2*...*E_n)= \Tr(E_n*E_1*...*E_{n-1})
(-1)^{k_n(k_1+\ldots+k_{n-1})},
\label{circ}
\end{eqnarray}
where $E_i$is the $k_i$ form in the base compact space. Then we
will take such equivalence  in  this sense  below,since the final result
is required
to integrate with one compact space. Additionally, we assume that the
physics fields disappear in infinite distance point,so {\bf R}$^D$ the
base space-time can be treated as {\bf S}$^D$ space. Throughout this papers
all calculation will be taken in {\bf S}$^D$ space. 
\par 
In path-integral formalism, the partition  function of a system involving
fermion fields and gauge fields is\cite{connes,JM,Landi}:
\begin{eqnarray}
Z= \int{\cal D}A_{\mu}^a{\cal D}\bar{\psi}{\cal D}
\psi e^{iS_G -\int\bar{\psi}\star
(i{\partial\!\!\!/}+{A\!\!\!/})\star\psi}, 
\end{eqnarray}
where 
\begin{eqnarray}\label{gaugeaction}
S_{G}= -\frac{1}{4}\int d^{D}x \ F_{\mu\nu}\left(x\right)
\star F^{\mu\nu}\left(x\right),
\end{eqnarray}
\begin{eqnarray}\label{gaugefields}
F_{\mu\nu}\left(x\right)&\equiv& \partial_{\mu}A_{\nu}\left(x\right)
-\partial_{\nu}A_{\mu}\left(x\right)+ig\big[A_{\mu}\left(x\right),
A_{\nu}\left(x\right)\big]_{\star}.
\end{eqnarray}
Set
\begin{eqnarray}\label{varder}
{D\!\!\!/} = {\gamma}^{\mu}\partial_{\mu}+{\gamma}^{\mu}[A_{\mu},\ \
]_{\star},
\end{eqnarray}
As usual,we first integrate the fermionic fields and treat the gauge
fields
as the background fields ,then finish out all quantization.  
\begin{eqnarray}
Z=\int{\cal D}A_{\mu}^{a}det{D\!\!\!/} e^{iS_G},~~~~~
det{D\!\!\!/}=\int{\cal D}\bar{\psi}{\cal D}\psi e^{-i\int\bar{\psi}\star
(i{\partial\!\!\!/}+{A\!\!\!/})\star\psi}.
\end{eqnarray}
From the ${det{D\!\!\!/}}$ one can define the WZW effective action
generated functional \cite{HOU}:
\begin{eqnarray}
W[A]=-ilndet{D\!\!\!/}.
\end{eqnarray}
Under the gauge transformation  $det{D\!\!\!/}$ changes into:
\begin{eqnarray}
Det\left({\partial\!\!\!/}+{A\!\!\!/}^g\right) =
e^{-i{\alpha}^1\left(A;g\right)}Det\left({\partial\!\!\!/}+{A\!\!\!/}\right),
\end{eqnarray}
where ${\alpha}^1\left(A;g\right)$ is assumed to exist, which is thought
reasonable from the existence of ${\alpha}^1\left(A;g\right)$  in
non-commutative two dimensions\cite{TWO,Ckr} and the true
consistent anomalies deduced from it in non-commutative four dimensions, 
and
then one can prove that it is the close-one-cochain of the
gauge group,which reveals the topology properties of the gauge group and
does not dependent
on the local property  of the gauge
group\cite{WZ,RS,BC,ZWZ,HOU,Guo}. Then we
can generalize
it in
non-commutative space, i.e.{\bf S}$^D$. 
\section{The WZW effective action in non-commutative four dimensions
space }
\noindent
The close-one-cochain of the gauge group in non-commutative four
dimensions
space,${\cal G}^4$,behaves as the first order topology obstacle,which
can be derived from the topology properties of the gauge group in
 higher by two dimensions,i.e. six-dimensions. So we now
start with the 
non-commutative {\bf S}$^6$. A-S index formula in
commutative geometry may not work in non-commutative space as
usual:
\begin{eqnarray}
Z = c_3
\int_{S_6}\Tr(F^3),~~~~~c_3={i^3\over{3!{\left(2\pi\right)}^3}}.
\label{F29}
\end{eqnarray}
We still call Z
Chen-Character though Z maybe is not an integer here \cite{Kmt}.
When $Z\neq 0$, ${{\Omega}^{-1}}_6\equiv{c_3\Tr(F\star F\star F)}$ is not
an exact form in {\bf S}$^6$,moreover the gauge potential $A$ cannot be
well-defined in {\bf S}$^6$. Insteadly,one can well-define
${{\Omega}^{-1}}_6(F)$  which is the close-zero-cochain in {\bf S}$^6$.
\par
We cover the manifold $M=S^6$ with two open sets:$U_0\cup U_1 = M$.
Each open set $U_j (j=0,1)$ is homomorphic to the six-dimensional plate
$D_6$ which is a trivial in topology . Then we can write
${{\Omega}^{-1}}_6(F)$ as an exact form in each trivial area $U_j$ due to
Poincare lemma:
\begin{eqnarray}
{{\Omega}^{-1}}_6(F=dA_j + A_j\star A_j)=d{{\Omega}^{0}}_5(A_j),
\end{eqnarray}
and we get easily 
\begin{eqnarray}
{{\Omega}^{0}}_5(A) & = & c_3\Tr(A\star dA\star dA + \frac 32  dA\star
A\star A\star A+ \frac 35  A\star A\star A\star A\star A) \nonumber \\
& = & c_3\Tr( A\star F\star F - \frac 12  A\star A\star A\star F + \frac 1{10}
 A\star A\star A\star A\star A).
\end{eqnarray}
${{\Omega}^{0}}_5(A_j)$ can only be defined in a trivial area
$U_j$ but in whole manifold
$S^6$.The${{\Omega}^{0}}_5(A_j)$ is a $\breve{C}ech$ zero-cochain
instead of $\breve{C}ech$ close-zero-cochain.In overlap  area
$U_{01}=U_0\cap U_1$,the difference of ${{\Omega}^{0}}_5(A_j)$ is
non-vanishing:
\begin{eqnarray}
\bigtriangleup {{\Omega}^{0}}_5(A_0,A_1)\equiv
{{\Omega}^{0}}_5(A_1)-{{\Omega}^{0}}_5(A_0).
\end{eqnarray}
Notice that $U_{01}$ and $S_5$ are homomorphic, so $U_{01}$ could be
condensed into $S_5$.Then 
\begin{eqnarray}
Z=\int_{S_5}\bigtriangleup {{\Omega}^{0}}_5(A_0,A_1).
\end{eqnarray}
Since the gauge potentials $\{ A_j\}$ in two areas $\{ U_j\}$ are
equivalent up to a
gauge transformation we can set:
\begin{eqnarray}
A_0 = A,A_1 = g^{-1}\star A\star g + g^{-1}\star dg.
\end{eqnarray}
Thus we have 
\begin{eqnarray}
\bigtriangleup {{\Omega}^{0}}_5(A,A^g) & = & 
\frac {c_3}2 d\Tr[A\star dA\star (dg\star g^{-1}) + dA\star A\star
(dg\star g^{-1}) \nonumber \\
&-&A\star (dg\star g^{-1})\star (dg\star
g^{-1})\star
(dg\star g^{-1})\nonumber \\ 
&-&\frac 12 A\star  (dg\star g^{-1})\star A\star  (dg\star
g^{-1}) + A\star  A\star  A\star (dg\star g^{-1})] \nonumber \\
&+&\frac {c_3}{10} 
\Tr[(dg\star g^{-1})\star(dg\star g^{-1})\star(dg\star
g^{-1})\star(dg\star
g^{-1})\star(dg\star g^{-1})]. 
\end{eqnarray}
 
\begin{eqnarray}
Z=\frac {c_3}{10}\int_{S_5}\Tr[(dg\star g^{-1})\star(dg\star
g^{-1})\star(dg\star g^{-1})\star(dg\star
g^{-1})\star(dg\star g^{-1})],
\end{eqnarray}

In fact,what we have done above is to map re strictly the differential
form $\wedge^*(S_6)$ in the manifold $S_6$ into each area $U_j$,and then
take the $\breve{C}ech$ difference of mapping,i.e. $\bigtriangleup$,which
leads to Mayer-Vietoris series when it works constantly:
\begin{eqnarray}
0\longrightarrow\wedge^*(S_6)\stackrel{r}{\longrightarrow}\wedge^*({D_6}^{(0)})\oplus\wedge^*({D_6}^{(1)})
\stackrel{\bigtriangleup}{\longrightarrow}\wedge^*(S_5)\longrightarrow 0,
\end{eqnarray}
where we should notice that ${\Omega^{-1}}_6(A_j)=d{\Omega^{0}}_5(A_j)$
is the $\breve{C}ech$ close-zero-cochain.Additionally, $\breve{C}ech$
difference $\bigtriangleup$ and differential operator d
commutate,which make\\ $\bigtriangleup {{\Omega}^{0}}_5(A_0,A_1)$ close
form. Then we can have $\bigtriangleup-d$ bi-complex.
\par 
In the same way we can cover $S_6$ with three,four,$\cdots$, areas
homomorphic with $D_6$ and operate  them with $d-\bigtriangleup$ orderly. 
Thus we have  the analogous series 
${{\Omega}^{1}}_4(A,A^{g_1}),{{\Omega}^{2}}_3(A,A^{g_1},A^{g_2})\cdots$
which satisfy:
\begin{eqnarray}
\bigtriangleup{{\Omega}^{k-1}}_{2n-k}(A^{g_0},A^{g_1},A^{g_2},\cdots,A^{g_k})
&\equiv&\sum_{i=0}^k {(-1)}^i
{{\Omega}^{k-1}}_{2n-k}(A^{g_0},A^{g_1},\cdots,{\breve{A}}^{g_i},\cdots,A^{g_k}),\\
\bigtriangleup{{\Omega}^{k-1}}_{2n-k}(A^{g_0},A^{g_1},A^{g_2},\cdots,A^{g_k})
 &=&
d{{\Omega}^{k}}_{2n-k-1}(A^{g_0},A^{g_1},A^{g_2},\cdots,A^{g_k}),\nonumber\\
& &(1\leq k\leq 2n,2n=6), 
\end{eqnarray}
where ${{\Omega}^{k}}_{2n-k-1}$ denotes the close-cochain density of gauge
group
space. Integrating 
with appropriate base space will give 
the close-cochains of the gauge group in the base space,that is  
\begin{eqnarray}
\alpha^1(A;g) = 2\pi\int_{S_4}{\Omega^1}_4(A,A^g)
\end{eqnarray}
This is the close-one-cochain of the gauge group,${\cal G}^4$ in the
$S_4($
i.e. physics $R^4)$.So we get our WZW effective action :  
\begin{eqnarray}
W[A,g] &=&2\pi\int_{S_4}{\Omega^1}_4(A,A^g)= 
 \alpha^1(\Gamma^1,S_4),\\
{\Omega^1}_4(A,A^g) & = &
\frac {c_3}2 \Tr[A\star dA\star (dg\star g^{-1}) + dA\star A\star
(dg\star g^{-1})\nonumber \\ 
&-&A\star (dg\star g^{-1})\star (dg\star g^{-1})\star
(dg\star g^{-1})\nonumber \\
&-&\frac 12 A\star  (dg\star g^{-1})\star A\star  (dg\star
g^{-1}) + A\star  A\star  A\star (dg\star g^{-1})] \nonumber \\
&+&\frac {c_3}{10}
d^{-1}\Tr[(dg\star g^{-1})\star(dg\star g^{-1})\star(dg\star 
g^{-1})\star(dg\star
g^{-1})\star(dg\star g^{-1})]{\label {omega}},
\end{eqnarray}
where $d^{-1}$ is a homotopic operator and only well-defined in a trivial
area in topology. Furthermore,we can argue the anomalies as expected if
we expand g around the unit $g = I + \Theta$: 
\begin{eqnarray}
W[A,I+\Theta]= -\frac {2\pi c_3 }2 \int_{S_4}\Tr(d\Theta \star A\star dA+
d\Theta \star dA\star A +
d\Theta \star A\star A\star A).
\end{eqnarray}
This is consistent with the results of L.Bonora and others
\cite{Bonora,Lang}.
\section {WZW Effective Action in Non-commutative high dimensions space }
\noindent
In principle ,the similar analysis can be generalized into high
dimensions. However,it will be very difficult to calculate the WZW
effective action.In this section ,we will take another way to get the 
close-one-cochain,even any order
close-cochain. We can calculate the close-cochains 
via the relations between the close-cochains of
connection space ${\cal U }= \{A(x)\}$ and gauge group,then the
${\Omega^1}_2n(A,A^g)$ and the  WZW effective action.
\par  At first we introduce the close-cochain of the connection space
${\cal U }= \{A(x)\}$
which is a poly manifold in an affine space of infinite dimensions.
So we have a linear insertion among k connections which have no
linear
relation :
\begin{eqnarray}
A_t = A_0 + {\sum^k}_{i=1}t_i\eta_i = A_\mu(x,t)dx^\mu,~~~
\eta_i = A_i - A_0,~~~0\leq{\sum^k}_{i=1}t_i\leq 1.  
\end{eqnarray}
Then  a connection is set
in $M\otimes{R(t)}^k$:
\begin{eqnarray}
{\cal A}=(A_\mu(x,t)dx^\mu;0) =  (A_t;0),
\end{eqnarray}
where ${R(t)}^k$ is a commutative linear space which commutates with base
space M too. The field strength is 
\begin{eqnarray}
{\cal F }= (d +\delta){\cal A}+ {\cal A}\star{\cal A}=
dA_t+{\sum^k}_{i=1}\delta t_i \eta_i + A_t\star A_t =
F_t +H,
\end{eqnarray}
where d is the derivative operator in $M$  and $\delta$ the derivative
operator with the arguments $\{t_i\}$ in arguments space ${R(t)}^k$,which
satisfy $d^2 = \delta^2 = 0,~~~~{(d+\delta)}^2=d\delta+\delta d=0$ and  
\begin{eqnarray}
F_t=dA_t + A_t \star A_t,~~~~
H={\sum^k}_{i=1}\delta t_i \eta_i \equiv \delta t \cdot \eta,
\end{eqnarray}
\par 
Expanding  the  trace of ${\cal F}^n$ in the power  of
H,one can get 
\begin{eqnarray}
\Tr({\cal F}^n)={\sum^n}_{k=1}{q^k}_{2n-k},\label{f}
\end{eqnarray}
with
\begin{eqnarray}
{q^k}_{2n-k} & = & \Tr[(H^k\star {F_t}^{n-k})+(H^{k-1}\star F_t 
\star H \star {F_t}^{n-k-1})+ \cdots + ({F_t}^{n-k}\star H^k)], 
\label{{q^k}_{2n-k}}
\end{eqnarray}
where there are $n!\over{k!(n-k)!}$ terms on the right hand side of
Equ.(\ref{{q^k}_{2n-k}}). On the other hand ,since 
\begin{eqnarray}
(\delta +d)\Tr ({\cal F}^n)=0.\label{trace}
\end{eqnarray}
multiplying $(d+\delta)$on the both sides of Equ.(\ref{f})  gives the 
descent equation:
\begin{eqnarray}
d{q^k}_{2n-k}=-\delta {q^{k-1}}_{2n-k+1},~~~k=1,2,\cdots,n\\
d{q^0}_{2n}=d\Tr({F_t}^n)=0,~~~~
\delta {q^n}_n(A_0,A_1,\cdots,A_n)=\delta \Tr(H^n)=0.
\end{eqnarray}
\par 
Integrate ${q^k}_{2n-k}$ with the k-order chain $\Gamma^k$ in the
connection space one can obtain the functional $Q_{2n-k}$ of $\Gamma^k$:
\begin{eqnarray}
\Gamma^k & = &   A_0+{\sum^k}_{i=1}t_i \eta_i
\equiv\Gamma^k(A_0,\cdots,A_k),
~~~~~~(0\leq{\sum^k}_{i=1}t_i\leq 1), \\
Q_{2n-k}(A_0,\cdots,A_k)&=&Q_{2n-k}(\Gamma^k)=
\int^{\Gamma^k}{q^k}_{2n-k}\nonumber \\
&=& {(-1)}^{{k(k-1)}\over 2}\int^{\bigtriangleup^k}{(\delta t)}^k
\Tr[(\eta_1 \star \eta_2 \star \cdots \eta_k \star {F_t}^{n-k})\nonumber 
\\
&+&(\eta_1 \star \eta_2 \star \cdots \eta_{k-1} \star F_t \star\eta_k 
{F_t}^{n-k-1})+ \cdots \nonumber \\ 
&+&({F_t}^{n-k} \star \eta_1 \star \eta_2
\star
\cdots \eta_k)],{\label {Q}}
\end{eqnarray}
where $Q_{2n-k}$ is the k-cochain in the connection space,called the
generalized Chen-Simons form $(Q$ series $)$. We can justify easily :
\begin{eqnarray}
dQ_{2n-k}(\Gamma^k)={(-1)}^{(k-1)}Q_{2n-k+1}(\partial \Gamma^k).
\end{eqnarray}
Define 
\begin{eqnarray}
Q_{2n-k+1}(\partial \Gamma^k)={\sum^k}_{i=1}{(-1)}^i
Q_{2n-k+1}({\Gamma_{(i)}}^{k-1})\equiv \bigtriangleup
Q_{2n-k+1}({\Gamma^{k}}),\nonumber \\
{\Gamma_{(i)}}^{k-1}\equiv {\Gamma}^{k-1}(A_0,\cdots ,{\breve
A}_i,\cdots,A_k),
\end{eqnarray}
we have 
\begin{eqnarray}
{\bigtriangleup}^2=0,
\end{eqnarray}
where $\bigtriangleup$ is called the coedge operator in the functional
space of connections. So we have our homology in the connection space.
In the following ,we pay our attention on  the expressions of Q series.
\par
According to  the following exact series between the homotopy groups via
injective and inclusion map: 
 \begin{eqnarray}
\pi_{n+1}{(\cal U)}\longrightarrow
\pi_{n+1}{({{\cal U}/ {\cal G}})}\longrightarrow\pi_{n}{(\cal G)}
\longrightarrow\pi_{n+1}{(\cal U)},
\end{eqnarray}
we can get  the cohomology group of gauge group  if we calculate the
cohomology group of connection space,i.e. Q series. In the expression
of $\Omega$ we can depart two
parts similar to Equ.(\ref{omega}): the Q series including the gauge
potential
A and R series including
only the pure gauge potential ${\it v}$,  
\begin{eqnarray} {\Omega^k}_{2n-k-1}(A,A^{g_1},\cdots ,A^{g_1 \cdots g_k})
=c_n(-1)^k Q_{2n-k-1}(0,A,{\it v}_1,\cdots ,{\it v}_k)
+{R^k}_{2n-k-1}(0,{\it v}_1,\cdots,{\it v}_k){\label {oq}},
\end{eqnarray}
where $c_n$ is defined similar to $c_3$ before  and 
\begin{eqnarray}
{\it v}_1=g_1 \star d{g_1}^{-1},~~{\it v}_2=(g_1 \star g_2)
d{(g_1 \star g_2)}^{-1},\cdots,
\end{eqnarray}
where Q series have explicit expressions in (\ref {Q}). In order to
calculate the R series we introduce the
generalized coedge operator  $\bar{\bigtriangleup}$:
\begin{eqnarray}
(-1)^{k-1}\bar{\bigtriangleup}Q_{2n-k}(0,A,{\it v}_1,\cdots,{\it v}_k)
={\bigtriangleup}Q_{2n-k}(0,A,{\it v}_1,\cdots ,{\it v}_k)-
Q_{2n-k}(0,{\it v}_1,\cdots,{\it v}_k).
\end{eqnarray}
Then we apply the coedge operator on the left hand side of the Equ.(\ref
{oq})
but on the right hand side  Q series must be applied with the generalized
coedge operator $\bar{\bigtriangleup}$:
\begin{eqnarray}
\bigtriangleup{\Omega^{k-1}}_{2n-k}(A,A^{g_1},\cdots,A^{g_1\cdots g_k})
&=&c_n(-1)^{k-1}\bar{\bigtriangleup} Q_{2n-k}(0,A,{\it v}_1,\cdots,{\it
v}_k)\nonumber\\
&+&\bigtriangleup{R^{k-1}}_{2n-k}(0,{\it v}_1,\cdots,{\it
v}_k),
{\label{doq1}}
\end{eqnarray}
Additionally, we act on both sides of Equ.(\ref {oq}) with the derivative
operator d:
\begin{eqnarray}
\bigtriangleup{\Omega^{k-1}}_{2n-k}(A,A^{g_1},\cdots,A^{g_1 \cdots g_k})
=c_n{\bigtriangleup} Q_{2n-k}(0,A,{\it v}_1,\cdots,{\it  v}_k)
+d{R^k}_{2n-k-1}(0,{\it v}_1,\cdots,{\it v}_k).{\label {doq2}}
\end{eqnarray}

Comparing Equ.(\ref {doq1}) and Equ.(\ref
{doq2})  we have the
formula satisfied by R series:
\begin{eqnarray}
d{R^k}_{2n-k-1}(0,{\it v}_1,\cdots,{\it v}_k)=c_n Q_{2n-k}(0,{\it v}_1,\cdots,{\it  v}_k)+ \bigtriangleup{R^{k-1}}_{2n-k}(0,{\it v}_1,\cdots ,{\it v}_k)
,{\label {R}}
\end{eqnarray}
i.e. the second term of Equ.(\ref {omega}) is a concrete example of above
formula. With this
formula we can calculate all order expressions of R series in the trivial
area by the homotopy operator $d^{-1}$, then we can have  all
explicit expressions of $\Omega$ if the explicit expressions of R series
are substituted in Equ.(\ref {omega}). Thus, we obtain the
$\Omega^1_{2n}$,i.e.
WZW effective Lagrangian and the WZW effective action when it is
integrated .
\section { Summary }
\noindent 
In this article we study the geometric property of the WZW effective
action and then
construct  the WZW effective action in non-commutative apace via
calculating
the close-one-cochain of the gauge group in non-commutative space . We
adopt two methods to construct the close-one-cochain in low dimensions
and high dimensions. In four dimensions  we cover the
six-dimensional $S^6$ with several trivial areas in topology
to derive the close-one-cochain in the non-commutative four dimensions
space. This method will become very difficult when it is generalized into
high dimensions space for it is too difficult to express a complex
expression with an exact form in a trivial area in topology of high
dimensions so we chose another way to
construct the close-cochains of gauge group. 
By calculating the cohomology group of the connection space
we obtained the close-cochains of gauge group in 
arbitrary even dimensions and further  calculated the
WZW effective action. 
Though our work stopped  here but we could   work more 
since we
have all close-cochains of the gauge group in the non-commutative high
dimensions space. The close cochains may be used to discuss some  physics
problems related to  
the topology properties of non-commutative high dimensions space, such as
the consistent anomalies \cite{Bonora,GM,Ard,Yang},Hamiltonian anomalies
and Jaccobi anomalies. 
\newpage

\centerline {\bf}

\end{document}